\documentclass[prepprint,,showpacs,amsmath,amssymb,superscriptaddress,aps,prl]{revtex4-1}
\usepackage{tabularx, booktabs}
\newcolumntype{Y}{>{\centering\arraybackslash}X}
\usepackage{graphicx}
\usepackage{dcolumn}
\usepackage{bm}

\begin{document}

\title{Lattice dynamics and the nature of structural transitions in organolead halide perovskites}

\author{Riccardo Comin}
\affiliation{Department of Electrical and Computer Engineering, University of Toronto. 10 King’s College Road, Toronto, Ontario, M5S 3G4, Canada}

\author{Michael K. Crawford}
\affiliation{DuPont Central Research and Development, Wilmington, Delaware 19880-0400, United States}

\author{Ayman Said}
\affiliation{Advanced Photon Source, Argonne National Lab, Argonne, Illinois 60439, United States}

\author{Norman Herron}
\affiliation{DuPont Central Research and Development, Wilmington, Delaware 19880-0400, United States}

\author{William E. Guise}
\affiliation{DuPont Central Research and Development, Wilmington, Delaware 19880-0400, United States}

\author{Xiaoping Wang}
\affiliation{Chemical and Engineering Materials Division, Oak Ridge National Laboratory, Oak Ridge, Tennessee 37831, United States}

\author{Pamela S. Whitfield}
\affiliation{Chemical and Engineering Materials Division, Oak Ridge National Laboratory, Oak Ridge, Tennessee 37831, United States}

\author{Ankit Jain}
\affiliation{Department of Electrical and Computer Engineering, University of Toronto. 10 King’s College Road, Toronto, Ontario, M5S 3G4, Canada}

\author{Xiwen Gong}
\affiliation{Department of Electrical and Computer Engineering, University of Toronto. 10 King’s College Road, Toronto, Ontario, M5S 3G4, Canada}

\author{Alan J. H. McGaughey}
\affiliation{Mechanical Engineering Department, Carnegie Mellon University, Pittsburgh, Pennsylvania 15213, United States}

\author{Edward H. Sargent}
\affiliation{Department of Electrical and Computer Engineering, University of Toronto. 10 King’s College Road, Toronto, Ontario, M5S 3G4, Canada}

\date{\today}

\begin{abstract}
Organolead halide perovskites are a family of hybrid organic-inorganic compounds whose remarkable optoelectronic properties have been under intensive scrutiny in recent years. Here we use inelastic X-ray scattering to study low-energy lattice excitations in single crystals of methylammonium lead iodide and bromide perovskites. Our findings confirm the displacive nature of the cubic-to-tetragonal phase transition, which is further shown, using neutron and x-ray diffraction, to be close to a tricritical point. Lastly, we detect quasistatic symmetry-breaking nanodomains persisting well into the high-temperature cubic phase, possibly stabilized by local defects. These findings reveal key structural properties of these materials, and also bear important implications for carrier dynamics across an extended temperature range relevant for photovoltaic applications. 
\end{abstract}

\pacs{Valid PACS appear here}
\maketitle

Structural phase transitions and tetragonal symmetry lowering have long been long studied in cubic halide and oxide perovskites \cite{Shirane_PbTiO3,fujii_CsPbCl3,Andrews_SrTiO3,Nicholls_KMnF3}. The nature of the phase transition is generally understood to be of the displacive type, involving the softening of a zone-edge mode (such as in CsPbCl$_3$, SrTiO$_3$, and KMnF$_3$) or a zone-center mode (such as in PbTiO$_3$) and the consequent frozen-in distortion of the lattice in the low-symmetry phase. This mechanism is accompanied by a change in symmetry which in reciprocal space is signalled by the emergence of a superlattice Bragg reflection at the wavevector where the phonon softening occurs.
In hybrid lead halide perovskites, a materials platform which has catalyzed considerable attention in the field of optoelectronics in recent years \cite{stranks_metal-halide_2015}, the crystal structure plays a fundamental role, exerting a direct feedback on the electronic structure and the bandgap \cite{noh_chemical_2013,mosconi_first-principles_2013,comin_structural_2015}. The resulting bandstructure and associated optical response are in turn crucial for photovoltaic and light emission applications, where spectral tunability is key for device performance and versatility. The influence of the structural changes on the performance of perovskite-based solar devices has been explored previously from both an experimental \cite{zhang_photovoltaic_2015} and a theoretical \cite{quarti_interplay_2014} viewpoint, across an extended temperature range relevant to photovoltaic functionality. Phenomenologically, it has been observed that the cubic-to-tetragonal phase transition has little effect on device performance, a finding that was later elucidated using \textit{ab initio} molecular dynamics simulations. This investigation revealed that on a sub-ps timescale the carriers still experience a slightly distorted (i.e. non-cubic) local environment, not dissimilar to that of the room-temperature tetragonal phase, thus lifting any discontinuity in the bandstructure and corresponding electrodynamics. A question then arises as to whether the local deviations from the cubic structure in the high-temperature phase are purely dynamic, or whether there is a static component, quenched by the presence of symmetry-breaking localized defects.
\begin{figure}[b!]
\includegraphics[width=0.65\linewidth]{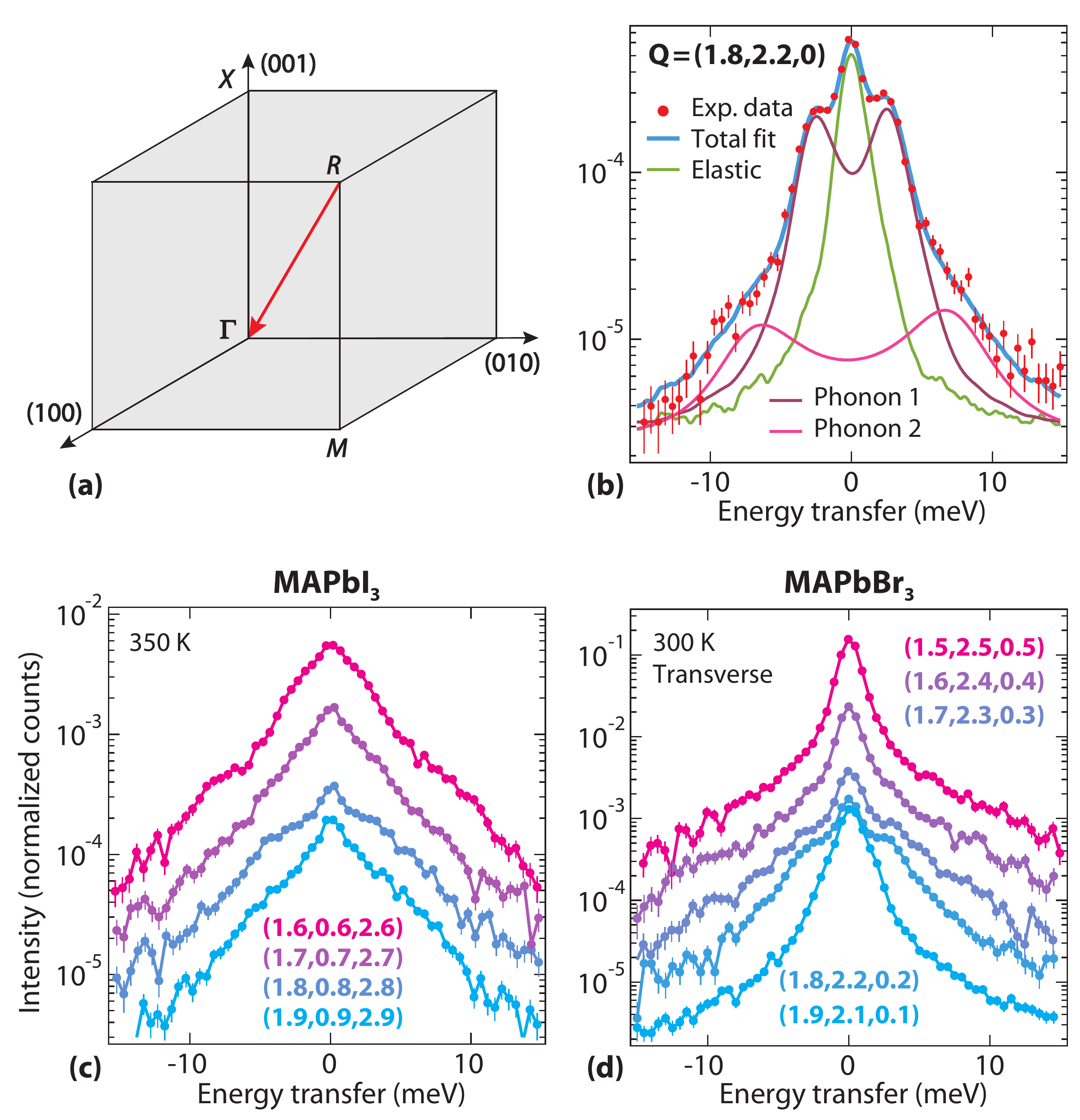}
\caption{(a) Reciprocal space coordinates for the cubic structure, with high-symmetry points highlighted. The red arrow defines the trajectory followed for the measurements shown in panels (c,d). (b) Single IXS scan from MAPbBr$_3$ showing a representative overall fit profile and its decomposition into elastic line and phonon sidebands. (c,d) A series of inelastic scans (scattered intensity vs. energy transfer) of MAPbI$_3$ (c) and MAPbBr$_3$ (d) at selected wavevectors along the path defined in (a).}
\label{Fig1}
\end{figure}

In order to assess such a scenario, in the present study we investigate the lattice dynamics of methylammonium (MA = CH$_3$NH$_3$) lead iodide (MAPbI$_3$) and bromide (MAPbBr$_3$) along the high-symmetry direction of the cubic phase of these compounds. MAPbI$_3$ crystallizes at room temperature in the tetragonal structure (space group $I 4 / m c m$) whereas MAPbBr$_3$ is found in the cubic phase (space group $P m \bar{3} m$) and both compounds undergo a cubic-to-tetragonal first-order structural transition at ${T}_{\mathrm{c}} \!=\! 327.5$\,K and $236.9$\,K, respectively \cite{poglitsch_dynamic_1987}. With the intent of clarifying the mechanistic details of the cubic-to-tetragonal phase transition, we use x-ray powder diffraction and neutron scattering to analyze the changes in lattice symmetry via the superlattice Bragg reflection, which serves as order parameter for the tetragonal structure. Additionally, we use inelastic X-ray scattering (IXS) to measure the dynamical structure factor as a function of momentum and energy in an extended portion of reciprocal space, and in the meV energy range. For these measurements, we synthesized crystals of MAPbI$_3$ and MAPbBr$_3$ using the inverse temperature crystallization method described in Ref.\,\onlinecite{kadro_facile_2015,saidaminov_high-quality_2015}. X-ray and neutron experiments have been performed at the superlattice reflection $\left( 3/2,1/2,1/2 \right)$, defined with respect to the cubic unit cell. In order to map out the complete phonon dispersion with IXS, we focused on reciprocal space directions connecting the Brillouin zone center point $\Gamma$ [with wavevector $ {\mathbf{Q}}_{\Gamma}\!=\! (H,K,L)$] to the zone edge high-symmetry points of the cubic phase, namely $X$ [$ {\mathbf{Q}}_{X}\!=\! (H+1/2,K,L)$], $M$ [$ {\mathbf{Q}}_{M}\!=\! (H+1/2,K+1/2,L)$], and $R$ [$ {\mathbf{Q}}_{R}\!=\! (H+1/2,K+1/2,L+1/2)$] \cite{HKL_note}. The location of these special points within the Brillouin zone of the high-temperature phase is shown in Fig.\,\ref{Fig1}(a). The red arrow defines the high-symmetry cut $R \rightarrow \Gamma$, for which a series of scans of the scattered intensity vs. energy transfer is shown at selected wavevectors for MAPbI$_3$ [Fig.\,\ref{Fig1}(c)] and MAPbBr$_3$ [Fig.\,\ref{Fig1}(d)]. All the IXS scans presented in this study have been measured in the cubic phase, at $T \!=\! 350$\,K and 300\,K for MAPbI$_3$ and MAPbBr$_3$, respectively. 
\begin{figure*}[t!]
\includegraphics[width=1\linewidth]{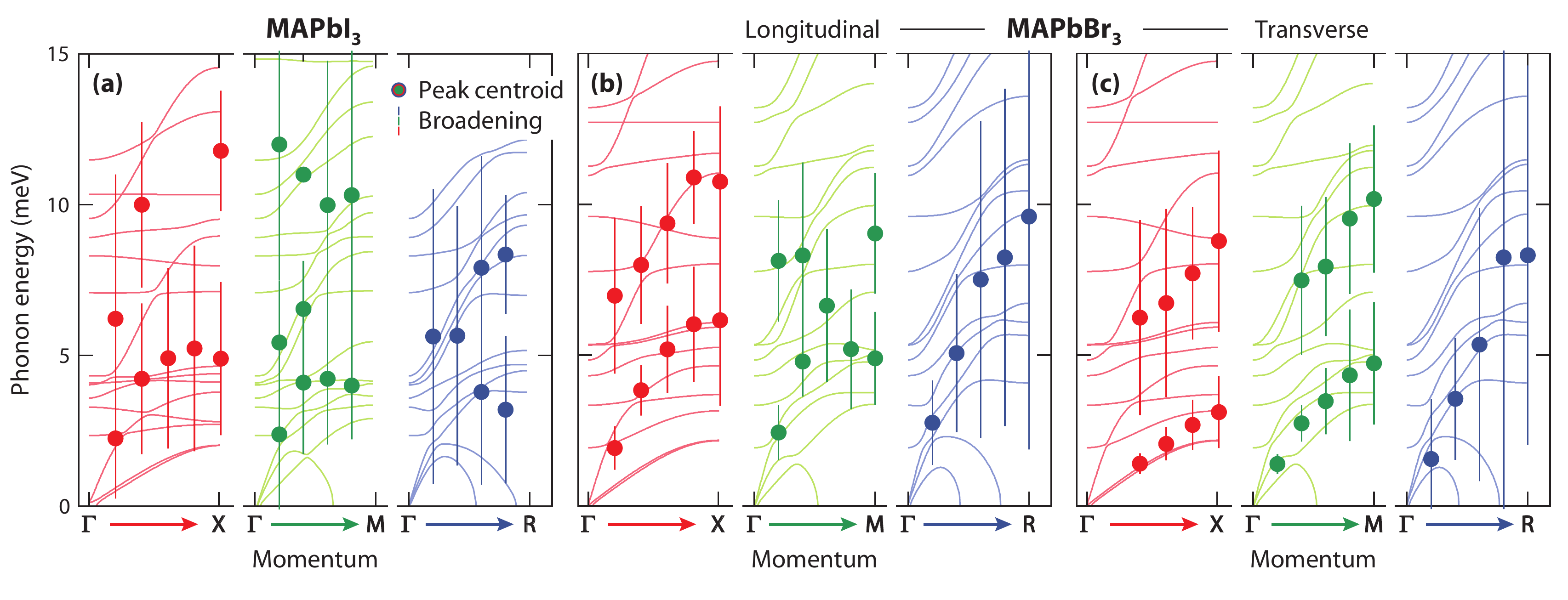}
\caption{(a) Experimental phonon dispersions (and calculated phonon bands overlaid) in MAPbI$_3$ along the three high-symmetry direction of the cubic phase, namely $\Gamma \rightarrow X$ (red, left), $\Gamma \rightarrow M$ (green, center), and $\Gamma \rightarrow R$ (blue, right). Longitudinal and transverse modes are mixed due to the scattering geometry adopted for MAPbI$_3$. (b,c) Experimental phonon dispersions (and calculated phonon bands overlaid) in MAPbBr$_3$ for longitudinal (b) and transverse (c) modes. Note: vertical bars reflect the experimental linewidth of phonon peaks and not the uncertainty on phonon frequency.}
\label{Fig2}
\end{figure*}

The IXS profiles have been analyzed using a functional form which incorporates a sharp ($\delta$-function) elastic line together with a damped-harmonic-oscillator (DHO) functional form \cite{Fak19971107}:
\begin{equation}
{I}_{\mathrm{DHO}} \left( \mathbf{Q}, \omega \right) = {I}_{\mathrm{el}} \cdot \delta \left( \omega \right) + {I}_{\mathrm{inel}} \cdot \left[ \dfrac{ {\Omega}_{\mathbf{Q}}^{2} {\Gamma}_{\mathbf{Q}} }{ {\left( {\omega}^{2} - {\Omega}_{\mathbf{Q}}^{2} \right)}^{2} + {\omega}^{2} {\Gamma}_{\mathbf{Q}}^{2} }  \right], 
\label{Eq_DHO}
\end{equation}
where $\omega$ is the photon angular frequency, ${\Omega}_{\mathbf{Q}}$ and ${\Gamma}_{\mathbf{Q}}$ are the momentum-dependent phonon frequency and damping (inverse lifetime), and ${I}_{\mathrm{el}}$ and  ${I}_{\mathrm{inel}}$ are the intensities of the elastic and inelastic terms, respectively. The final model function applied to the analysis of the experimental data is obtained after multiplication of Eq.\,\ref{Eq_DHO} by the Bose factor $\left[ n \left( \omega \right) +1 \right]$ to account for the temperature-dependent phonon population [$ n \left( \omega \right)$ being the Bose-Einstein function] and subsequent convolution with the instrument resolution function (see also Supplemental Material). An example of the fit profile based on this model function is shown in Fig.\,\ref{Fig1}(b), where two sets of phonon sidebands can be resolved, together with the central elastic line.
\begin{figure}[t!]
\includegraphics[width=0.65\linewidth]{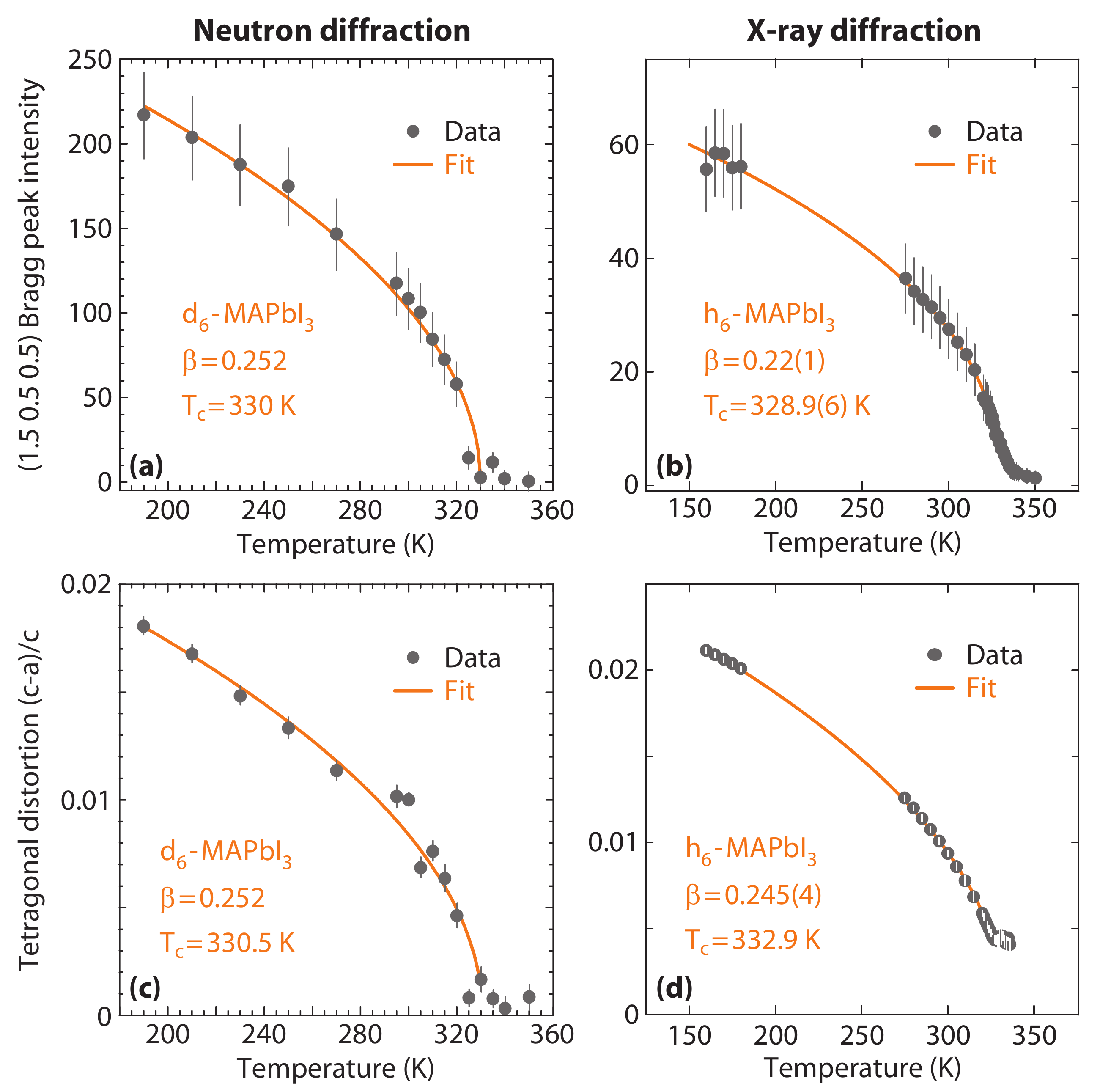}
\caption{(a,c) Cubic (3/2 1/2 1/2) superlattice Bragg intensity (a) and tetragonal distortion (c) for single crystal d$_6$-MAPbI$_3$ measured by neutron diffraction on TOPAZ. (b,d) Cubic (3/2 1/2 1/2) superlattice Bragg intensity and tetragonal distortion for polycrystalline h$_6$-MAPbI$_3$ measured by X-ray powder diffraction on DND-CAT. Error bars in panel (d) are multiplied by a factor 3 for clarity.}
\label{Fig3}
\end{figure}
\begin{figure}[t!]
\includegraphics[width=0.65\linewidth]{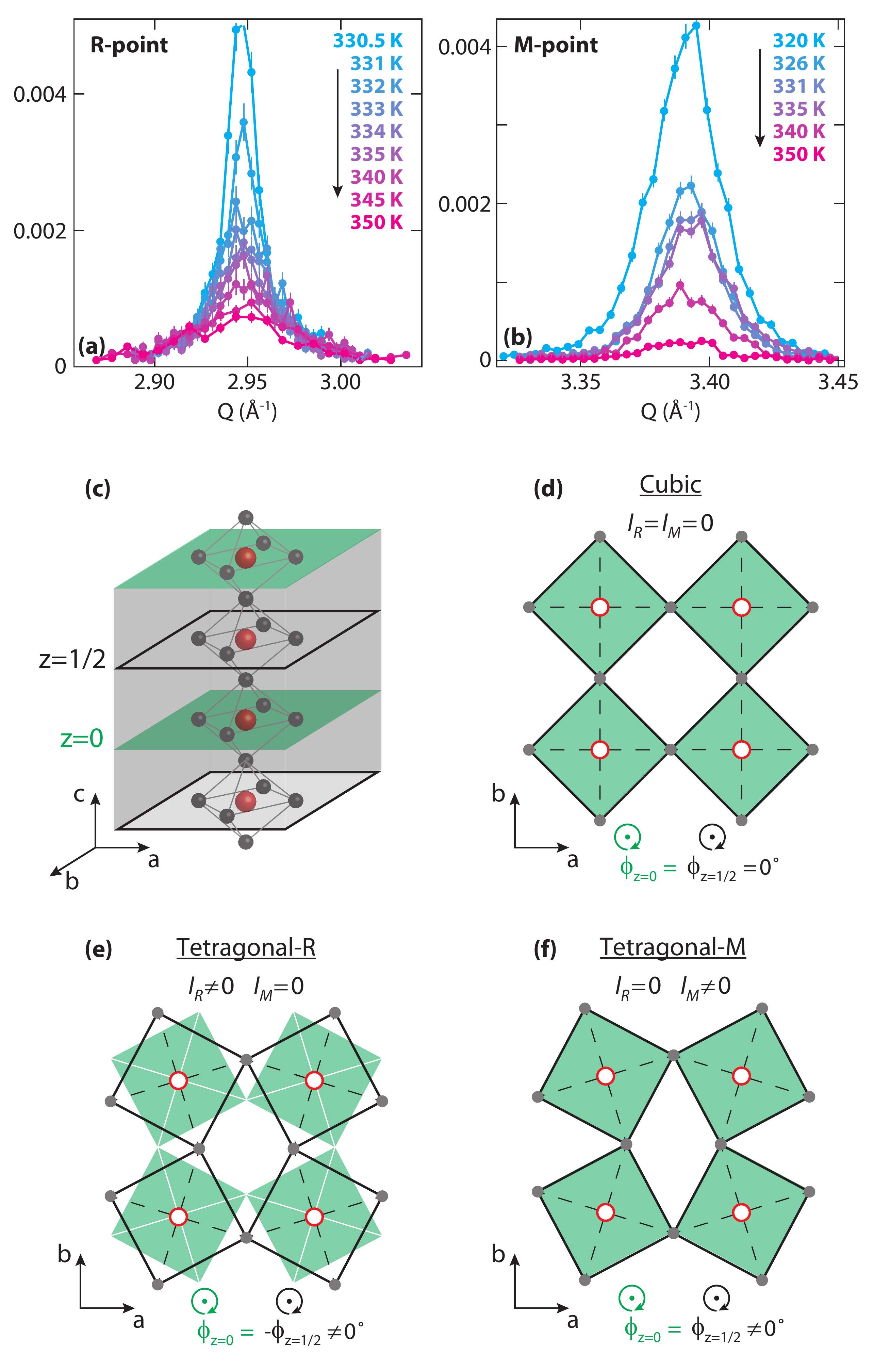}
\caption{(a,b) Temperature-dependent momentum scans of MAPbI$_3$ at the $R$- (a) and $M$-point (b) above the cubic-to-tetragonal phase transition temperature. (c-f) Side view of the crystal structure (c), and top view along c-axis highlighting the different octahedral rotation motifs for the cubic (undistorted) phase (d), tetragonal-$R$ phase (e), and tetragonal-$M$ phase (f). Note that the two tetragonal phases have different unit cells (the one of tetragonal-$R$ structure is doubled along the c-axis) and their order parameters are represented by the Bragg intensity at $R$- and $M$-point, respectively.}
\label{Fig4}
\end{figure}

By applying this analysis to a complete dataset of IXS experimental profiles, we have extracted the phonon dispersions in MAPbI$_3$ [Fig.\,\ref{Fig2}(a)] and MAPbBr$_3$ for longitudinal [Fig.\,\ref{Fig2}(b)] and transverse modes [Fig.\,\ref{Fig2}(c)], along the $\Gamma \rightarrow X$ (red), $\Gamma \rightarrow M$ (green), and $\Gamma \rightarrow R$ (blue) directions. From numerical fits of the experimental data to Eq.\,\ref{Eq_DHO}, we have derived both the phonon frequency (markers) and broadening (vertical bars). The number of branches that we could resolve depends on the counting statistics and varies with the wavevector, but in all cases is limited by the range of energy transfer ($\pm 15$\,meV). All the modes detected in our measurements corresponds to low-energy excitations of the corner-sharing PbX$_3$ (X = I, Br) network, while the methylammonium internal vibrations generally appear at higher energies, although the translational and librational modes are found in the 10-20 meV energy range \cite{quarti_raman_2014,brivio_lattice_2015}. While it is possible that multiple modes are lumped together within individual fitted peaks, the broad energy linewidths observed at various momenta might also indicate short phonon lifetimes, as recently observed using Raman spectroscopy \cite{leguy_dynamic_2016}. The experimental phonon frequencies are compared to the theoretical vibrational bandstructure calculated using density functional theory in the frozen-phonon approximation [dispersions are overlaid in Figs.\,\ref{Fig2} and are in close agreement with those reported in Ref.\,\onlinecite{brivio_lattice_2015}]. In addition, the highest zone-center optical mode in MAPbI$_3$ is found at 7-7.5\,meV, consistent with recent findings from a Raman/IR study \cite{perez-osorio_vibrational_2015}. Additionally, we note that the phonons become overdamped at the $R$-point (this is particularly evident for MAPbBr$_3$), where numerical calculations predict an instability as signalled by the imaginary mode frequency, consistent with a displacive structural transition triggered by phonon condensation at the zone-corner. A closer inspection of the longitudinal (LA) and transverse acoustic (TA) phonon dispersions in the vicinity of the $\Gamma$-point (along the $\Gamma \rightarrow X$ direction) yields an estimate of the sound speed, which is a relevant quantity with a direct impact on the carrier mobility $\mu$ via electron-phonon coupling \cite{takagi_universality_1994}. This analysis indicates a longitudinal (shear) sound speed of ${v}_{l} \!=\! 2714 \pm 650$\,m/s (${v}_{s} \!=\! 990 \pm 150$\,m/s) in MAPbBr$_3$, while in MAPbI$_3$ this procedure is hindered by the inability to separate the LA and TA branches, which are mixed in due to the experimental geometry. However, from the calculated phonon bands we obtain ${v}_{l} \!\sim\! 3270$\,m/s and ${v}_{s} \!\sim\! 970$\,m/s. The sound speed measured using IXS slightly underestimates the sound speed in the hydrodynamic regime ($\sim 1200$\,m/s) as measured using picosecond acoustic dynamics \cite{guo_structural_2016}. Further, these values suggest that the electron-phonon scattering rate ${\tau}^{-1}$ from acoustic modes, which is inversely proportional to the sound speed (${\tau}^{-1} \!\propto\! {v}^{2}$, see Ref.\,\onlinecite{filippetti_low_2016}), is slightly higher in MAPbI$_3$ and MAPbBr$_3$. However, it was recently pointed out that scattering of electrons from optical phonons is the dominant mechanism behind electron transport in these materials \cite{wright_electron-phonon_2016,leguy_dynamic_2016,filippetti_low_2016}, likely explaining the long carrier lifetimes and diffusion lengths observed in high-quality crystals of hybrid halide perovskites \cite{shi_low_2015}.

In the second part of our study we examined the momentum fingerprints of the cubic-to-tetragonal structural transition and how the lowering of the lattice symmetry is unfolded in reciprocal space. 

The nature of the cubic-tetragonal phase transition was also investigated using single crystal neutron diffraction at the ORNL Spallation Neutron Source's TOPAZ instrument \cite{Schultz_TOPAZ} for a deuterated MAPbI$_3$ crystal (d$_6$-MAPbI$_3$) and synchrotron X-ray powder diffraction for a fully hydrogenated polycrystalline sample (h$_6$-MAPbI$_3$). The intensity of the (3/2 1/2 1/2) superlattice Bragg peak and the tetragonal distortion were measured to evaluate the critical behavior of the phase transition, and the results are shown in Fig.\,\ref{Fig3}. Each set of data was fit to a power law, $I \sim {\left( {T}_{\mathrm{c}} - T \right)}^{2 \beta} $, where the factor of two in the exponent accounts for the fact that the superlattice intensity will scale as the square of the order parameter for the phase transition, where the order parameter is the rotation angle of the PbI$_6$ octahedra \cite{kawamura_structural_2002}. Similarly, the tetragonal strain is a secondary order parameter which is also expected to scale as the square of the order parameter for the phase transition \cite{Slonczewski_1970}.
The values of $\beta$ vary from 0.249 to 0.28, where the mean-field result for a second-order phase transition is $\beta \!=\! 0.5$. The measured values are close to that expected for a tricritical phase transition, $\beta \!=\! 0.25$, which occurs at the intersection of lines of first-order and second-order phase transitions \cite{Landau_phase_transitions_Toledano}. However, the presence of a distinct temperature range of cubic and tetragonal phase coexistence \cite{Whitfield_2016}, and latent heat in heat capacity measurements \cite{OnodaYamamuro1990}, are only consistent with a first-order transition, suggesting that these materials are still on the first-order side of the tricritical point.

Furthermore, we have performed IXS scans in momentum space across the $R$- and $M$-points in MAPbI$_3$, as a function of temperature above the phase transition [Figs.\,\ref{Fig4}(a) and (b)]. More specifically, we have scanned reciprocal space by sweeping the detector (and therefore the scattering) angle while in energy-resolved mode, i.e. collecting only the elastically scattered light at $\omega \!=\! 0$ and at low momentum resolution (0.01 \AA${}^{-1}$) to maximize the sensitivity to weak scattering signals. Most importantly, the high energy resolution allows us to separate the elastic (quasistatic) component from the thermal diffuse scattering, which is inelastic in nature and cannot be avoided in energy-integrated measurements such as neutron or X-ray diffraction, discussed above.
At variance with previous reports \cite{weller_complete_2015}, we find that the (elastic) scattered intensity at the $R$-point does not vanish above ${T}_{\mathrm{c}}$, but persists up to 20\,K away from the transition temperature, as shown in Fig.\,\ref{Fig3}(a) (a similar behavior is found in MAPbBr$_3$, see Supplemental Material). In MAPbI$_3$ we also detect weaker reflections at the $M$-point [Fig.\,\ref{Fig3}(b)], which suggests that, alongside the dominant lattice symmetry and its representative Bragg signatures, there are small regions with short-range-ordered puddles locally breaking cubic symmetry. Since our data at $\omega \!=\! 0$ only reflect the static scattering, they demonstrate the existence of symmetry-breaking nanopuddles well above the thermodynamic structural transition temperature, in analogy with the phenomenology of the central peak in SrTiO$_3$ \cite{holt_dynamic_2007,hong_central_2008}. The associated temperature evolution of the forbidden superlattice reflection intensities (which in the cubic phase are $\sim$ 2 orders of magnitude weaker than the allowed $R$-point Bragg reflection below ${T}_{\mathrm{c}}$) and correlation lengths (see Supplemental Material) suggests that these nanoislands initially nucleate around defects that break cubic symmetry and progressively grow in size coalescing into long-range ordered structures near and below the phase transition. The formation of these cubic-symmetry-breaking domains therefore precedes the complete renormalization of the phonon frequency and the concomitant condensation of the phonon mode into a static distortion. These two phenomena are \textit{a priori} independent, and while the phonon softening mechanism is driven by a native instability of the cubic lattice, the short-ranged nanodomains likely reflect the nature of the symmetry-breaking defects, previously investigated \cite{Buin_2014,Buin_2015}. This likely explains why we observe scattering intensity also at the $M$-point, where no frozen-in phonon modes should be expected altogether.

In conclusion, the present study confirms the existence of a lattice instability at the $R$-point, and elucidates the tricritical nature of the cubic-to-tetragonal phase transition. In addition, we uncover the presence of static short-ranged order well above the structural transition and in the temperature range relevant for photovoltaic operation.

\vspace{2mm}
\begin{center}
\noindent {\bf Acknowledgements}
\end{center}
A portion of this research at Oak Ridge National Laboratory’s Spallation Neutron Source was sponsored by the Scientific User Facilities Division, Office of Basic Energy Sciences, U. S. Department of Energy. Portions of this work were performed at the DuPont-Northwestern-Dow Collaborative Access Team (DND-CAT) located at Sector 5 of the Advanced Photon Source (APS). DND-CAT is supported by Northwestern University, E.I. DuPont de Nemours \& Co., and The Dow Chemical Company. This research used resources of the Advanced Photon Source, a U.S. Department of Energy (DOE) Office of Science User Facility operated for the DOE Office of Science by Argonne National Laboratory under Contract No. DE-AC02-06CH11357.

\clearpage

\renewcommand{\bibname}{References}
\setcounter{figure}{0}
\renewcommand{\thefigure}{S\arabic{figure}}

\begin{center}
\Large{Supplemental Material}

\vspace{6mm}

\large{\textbf{Lattice dynamics and the nature of structural transitions in organolead halide perovskites}}

\vspace{4mm}

\normalsize{Riccardo Comin,$^{1}$ Michael K. Crawford,$^{2}$ Ayman Said,$^{3}$ Norman Herron,$^{2}$ William E. Guise,$^{2}$ Xiaoping Wang,$^{4}$ Pamela S. Whitfield,$^{4}$ Ankit Jain,$^{1}$ Xiwen Gong,$^{1}$ Alan J. H. McGaughey,$^{5}$ Edward H. Sargent$^{1}$}

\vspace{4mm}

\vspace{6mm}

\author{Riccardo Comin}
\author{Michael K. Crawford}
\author{Ayman Said}
\author{Norman Herron}
\author{William E. Guise}
\author{Xiaoping Wang}
\author{Pamela S. Whitfield}
\author{Ankit Jain}
\author{Xiwen Gong}
\author{Alan J. H. McGaughey}
\author{Edward H. Sargent}
\normalsize{$^{1}$Department of Electrical and Computer Engineering, University of Toronto. 10 King’s College Road, Toronto, Ontario, M5S 3G4, Canada}\\
\normalsize{$^{2}$DuPont Central Research and Development, Wilmington, Delaware 19880-0400, United States}\\
\normalsize{$^{3}$Advanced Photon Source, Argonne National Lab, Argonne, Illinois 60439, United States}\\
\normalsize{$^{4}$Chemical and Engineering Materials Division, Oak Ridge National Laboratory, Oak Ridge, Tennessee 37831, United States}\\
\normalsize{$^{5}$Mechanical Engineering Department, Carnegie Mellon University, Pittsburgh, Pennsylvania 15213, United States}\\

\vspace{4mm}

\end{center}


\section{Methods}%

\subsection{Phonon calculations}

The phonon vibration frequencies are obtained by diagonalizing the dynamical matrix which was calculated using the frozen phonon approach. The second order force constants, which are required as an input for dynamical matrix, are obtained from the finite difference of Hellmann-Feynman forces obtained on a $2 \times 2 \times 2$ supercell consisting of 96 atoms from density functional theory package VASP. We employed electronic wavevector grid of $3 \times 3 \times 3$  and planewave energy cutoff of 520\,eV with PBEsol exchange correlation based projected augmented wave pseudopotential.

\subsection{Dynamical structure factor calculations}

We have explored the influence of a dynamical lattice on the measured x-ray scattering signal. The link between theory and experiments is provided by the dynamic structure factor (DSF) $ S \left( \mathbf{Q}, \omega \right) $, which can be evaluated from the output of molecular dynamics (MD) calculations of the atomic motion in a supercell of the compound CH$_3$NH$_3$PbI$_3$, and directly compared to the IXS data, whose associated observable is proportional to the DSF.

\begin{figure}[t!]
\includegraphics[width=0.65\linewidth]{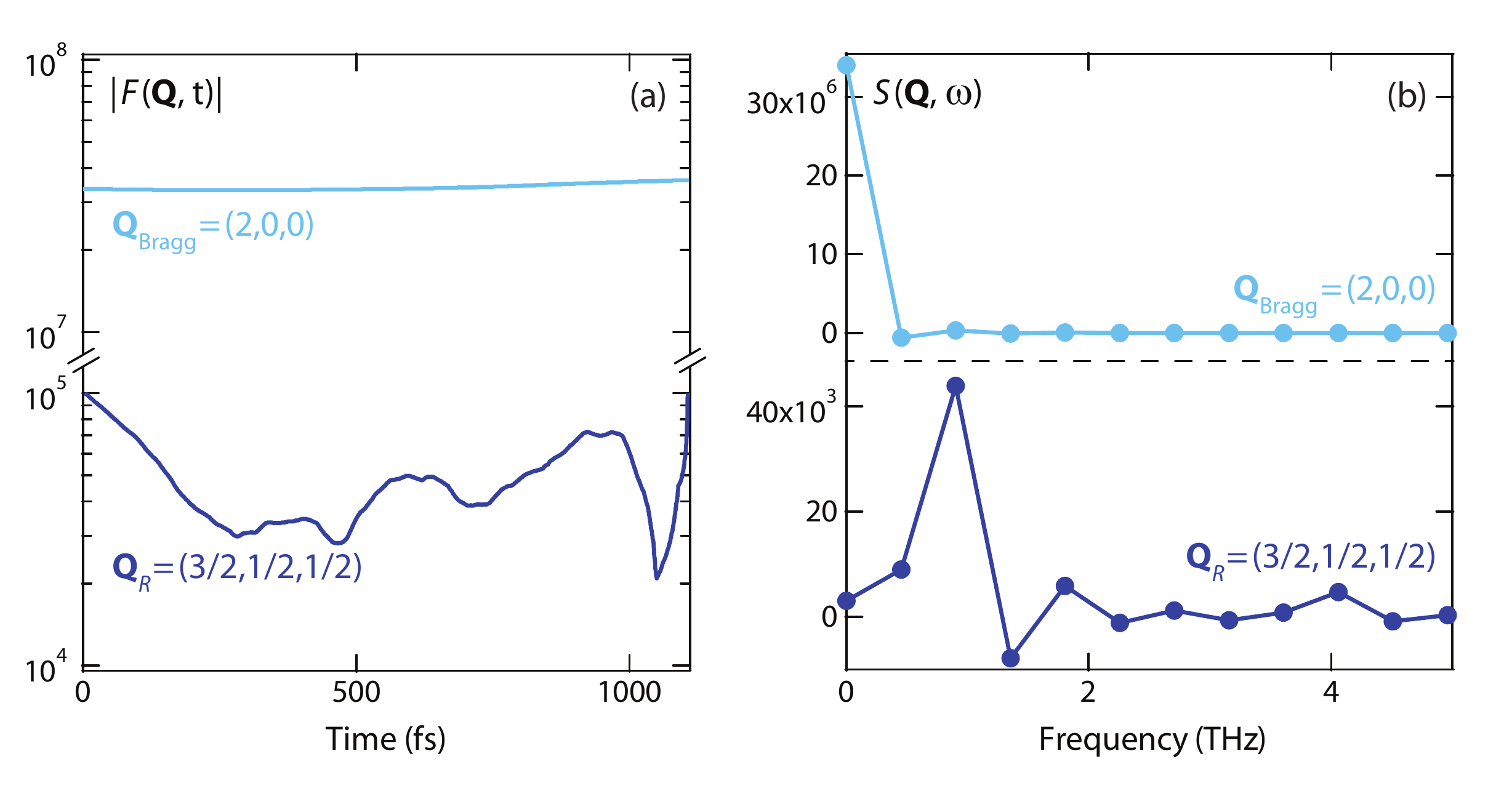}
\caption{(a) Time traces of the intermediate scattering function $F \left( \mathbf{Q}, t \right)$ at a Bragg reciprocal lattice vector ${\mathbf{Q}}_{\mathrm{Bragg}} \!=\! \left(2,0,0 \right)$ and at the $R$-point ${\mathbf{Q}}_{R} \!=\! \left( 3/2, 1/2, 1/2 \right)$. Note the broken vertical axis and the logarithmic scale. (b) Corresponding traces of the dynamical structure factor $S \left(\mathbf{Q}, \omega \right)$ as a function of frequency $\omega$, for the same momentum vectors.}
\label{SI_ISF}
\end{figure}

We performed our MD simulations in NVT ensemble using a Verlet algorithm to integrate Newton's equations of motion. We employed a $3 \times 3 \times 3$ supercell consisting of 324 atoms with a Gamma-point electronic wavevector grid and a time-step of 0.5 fs. We allowed the system to equilibrate for 10,000 timesteps before collecting trajectory data for subsequent 8000 timesteps at a temperature of 300\,K. All of our calculations are performed using a density functional theory package VASP. The output of the MD calculations consists in a full set of atomic positions as a function of time step, ${\mathbf{R}}_{i} \left( t \right) $ (the index $i$ labels the $i$-th atom in the selected supercell). From these time-dependent atomic coordinates, we can calculate the \textit{intermediate scattering function} $F \left( \mathbf{Q}, t \right) \!=\! \left\langle {\rho}_{\mathbf{-Q}} \cdot {\rho}_{\mathbf{Q}} \left( t \right) \right\rangle $, which encodes the time-dependent density-density correlations in reciprocal space. In the above definition, ${\rho}_{\mathbf{Q}}$ is the Fourier transform of the atomic density and can be expanded as ${\rho}_{\mathbf{Q}} \!=\! \int d \mathbf{r} {e}^{-i \mathbf{Q} \cdot \mathbf{r}} \rho \left( \mathbf{r} \right) \!=\! {\sum}_{i} {e}^{-i \mathbf{Q} \cdot {\mathbf{R}}_{i}} $, while $\left\langle \ldots \right\rangle $ is a thermodynamic ensemble average. Numerically, the calculation of $F \left( \mathbf{Q}, t \right)$ is performed using the atomic coordinates and replacing the thermodynamic average with a time average ${\left\langle \ldots \right\rangle}_{t} $:
\begin{equation}
F \left( \mathbf{Q}, t \right) = {\left\langle {\sum}_{i,j} {e}^{-i \mathbf{Q} \cdot \left[ {\mathbf{R}}_{j} \left( t \right) - {\mathbf{R}}_{i} \right]} \right\rangle}_{t} = \dfrac{1}{N(t)} \cdot {\sum}_{t'=0,N(t)} {\sum}_{i,j} {e}^{-i \mathbf{Q} \cdot \left[ {\mathbf{R}}_{j} \left( t + t' \right) - {\mathbf{R}}_{i} \left( t' \right) \right]}
\end{equation}
At each time step $t$ the intermediate scattering function is evaluated on a momentum grid via a double summation over atomic coordinate indexes $i$ and $j$, and by performing averaging over the maximal time window available $\left[ 0, N(t) \right]$ which depends on the time coordinate $t$. In Fig.\,\ref{SI_ISF}(a) we display representative traces of the squared modulus of $F \left( \mathbf{Q}, t \right)$ (which is, in general, a complex quantity) vs. time at a Bragg reflection and at the $R$-point. Inspection of these traces immediately shows the large amplitude and static nature of the Bragg scattering, whose intensity is concentrated at $\omega \!=\! 0$. In contrast, the scattering function at the $R$-point is suppressed by almost 3 orders of magnitude and it oscillates slowly, on a timescale of hundreds of femtoseconds.
The intermediate scattering function can be Fourier-transformed to obtain the DSF: $S \left(\mathbf{Q}, \omega \right) \!=\! \int d t {e}^{-i \omega t} F \left( \mathbf{Q}, t \right) $. Representative traces of the DSF at the same momenta as above are plotted in Fig.\,\ref{SI_ISF}(b). Note again the large difference in the scattering amplitude, as well as the vanishing of any DC ($\omega \!=\! 0$) component for $S \left( {\mathbf{Q}}_{R}, \omega \right)$, which signals the absence of any elastic scattering at the $R$-point. We find however a clear peak in $S \left( {\mathbf{Q}}_{R}, \omega \right)$ at $\omega \sim 1$\,THz ($\sim 4$\,meV), which likely reflects the presence of slow oscillations in $F \left( {\mathbf{Q}}_{R}, t \right)$. Interestingly, this energy scale matches the position of the side humps observed in the quasielastic scattering measurements performed in the cubic phase (see also Fig.\,\ref{SI_Fig3} and \ref{SI_Fig4} and corresponding discussion) . From these data, we can conclude that molecular dynamics calculations seem to indicate that the only scattering occurring at the $R$-point is dynamical in nature.
\begin{figure}[b!]
\includegraphics[width=1\linewidth]{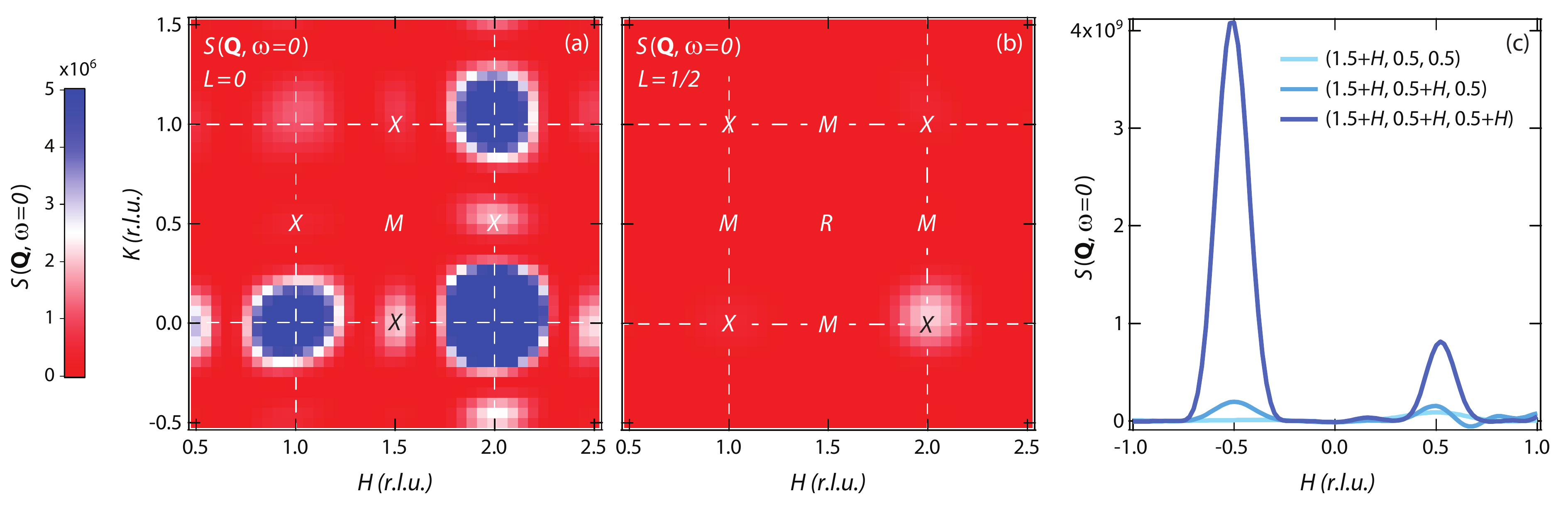}
\caption{(a,b) Dynamic structure factor $S \left(\mathbf{Q}, \omega \right)$ at $\omega \!=\! 0$, projected onto the $H-K$ plane at $L \!=\! 0$ (a) and $L \!=\! 1/2$ (b). (c) One-dimensional momentum cuts of $S \left(\mathbf{Q}, \omega \right)$ across the $R$-point ${\mathbf{Q}}_{R} \!=\! \left( 3/2, 1/2, 1/2 \right)$ along the three high-symmetry directions of the cubic lattice: $(100)$ (light blue); $(110)$ (blue); $(111)$ (dark blue).}
\label{SI_DSF}
\end{figure}
In Figure\,\ref{SI_DSF}(a,b) we additionally plot the momentum-resolved DSF at zero energy ($\omega \!=\! 0$) and in the $H-K$ plane (reciprocal lattice axes are indexed in the cubic structure) at $L \!=\! 0$ and $L \!=\! 1/2$, respectively. The $H-K$ momentum grid covers the range $0.5 < H < 2.5$ and $-0.5 < K < 1.5$; this is chosen so that the $L \!=\! 1/2$ slice is centered at the R-point ${\mathbf{Q}}_{R} \!=\! \left( 3/2, 1/2, 1/2 \right)$. As it is clearly visible, the strongest feature is the Bragg peak at $\left( 2, 0, 0 \right)$, which is also seen in the $L \!=\! 1/2$ slice due to tailing intensity along the $L$ axis (likely a consequence of the finite size of the supercell). We similarly observe large intensity at the $\left( 1, 0, 0 \right)$ and $\left( 2, 1, 0 \right)$ Bragg vectors. Notably, we find no intensity at the $M$- or $R$-points, whereas minor satellite peaks can be seen at some of the $X$-points [particularly those located around the strongest Bragg peak $(200)$], again likely due to fringe oscillations in momentum space that are induced by the finite size of the real space supercell. One-dimensional line profiles along the high-symmetry directions $(100)$ ($\Gamma \rightarrow X$), $(110)$ ($\Gamma \rightarrow M$), and $(111)$ ($\Gamma \rightarrow R$) are further displayed in Fig.\,\ref{SI_DSF}(c); the center point of all line cuts is the $R$-point ${\mathbf{Q}}_{R} \!=\! \left( 3/2, 1/2, 1/2 \right)$. The $(111)$ cut crosses the two Bragg vectors $\left( 1, 0, 0 \right)$ (left peak) and $\left( 2, 1, 1 \right)$ (right peak), whereas the other two cuts do not intersect any integer peaks, however a small bump can be seen in the $(110)$ cut at a wavevector $\left( 1, 0, 0.5 \right)$ which is explained as one of the fringe oscillations arising from Fourier-transforming a finite-sized support in real space. Most importantly, no intensity can be seen around the $M$- or $R$-points in the $(100)$ cut, which would be located at, respectively, $H \!=\! \pm 0.5$ and $H \!=\! 0$ on the horizontal scale. This is consistent with the color maps in Fig.\,\ref{SI_DSF}(a,b) and further confirms that there is no elastic scattering at $M$- or $R$-points to be expected from MD calculations performed in the cubic symmetry.

\subsection{Inelastic x-ray scattering measurements and analysis}

\begin{figure}[h!]
\includegraphics[width=0.35\linewidth]{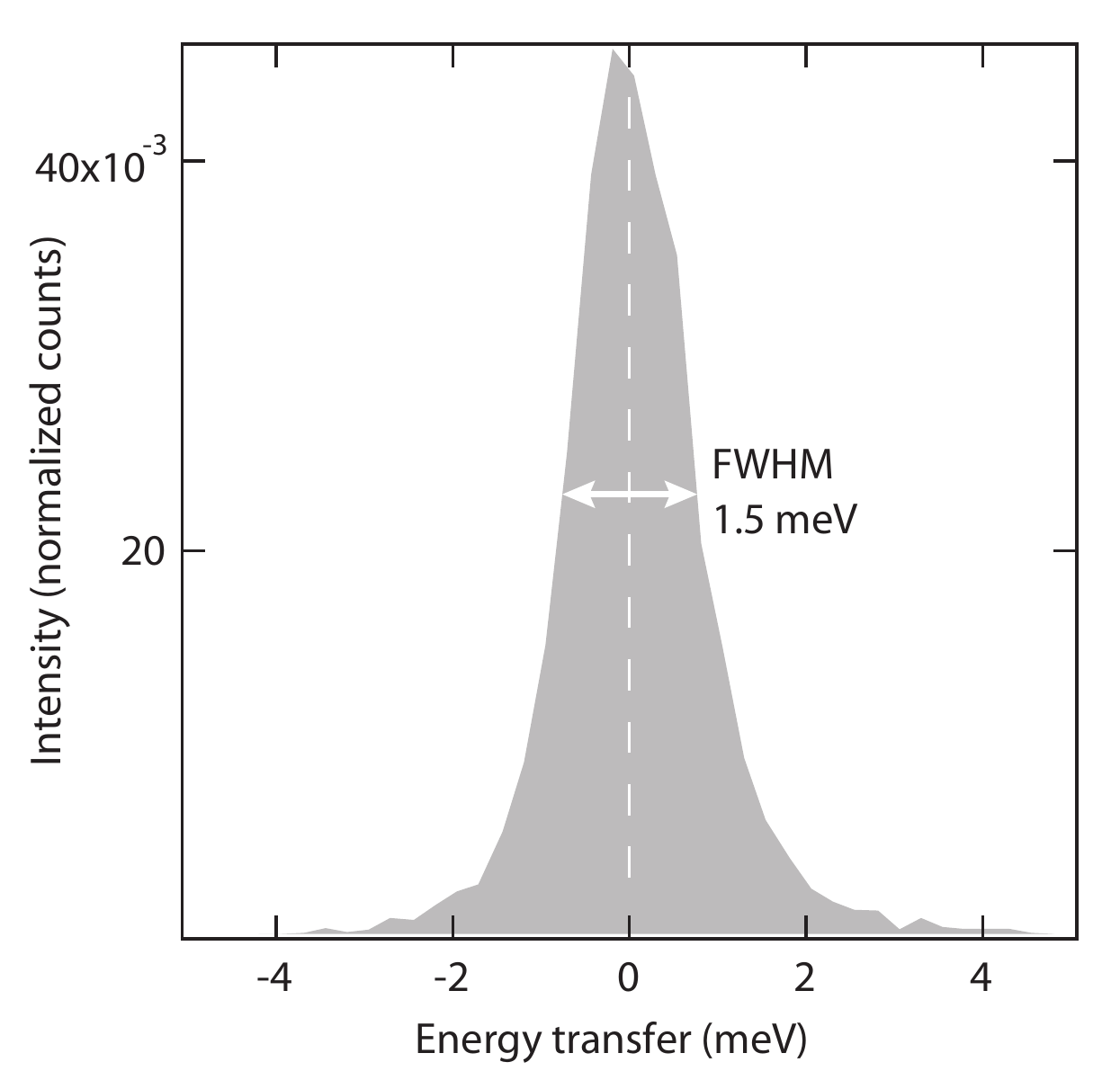}
\caption{Experimental IXS spectral resolution profile, measured from the elastic scattering off a Plexiglass\textsuperscript{\textregistered} slab, with a FWHM (full-width-at-half-maximum) of around 1.5\,meV.}
\label{SI_Resolution}
\end{figure}

We performed inelastic x-ray scattering (IXS) measurements at Sector 30 of the Advanced Photon Source, on a four-circle diffractometer with a combined momentum resolution of 0.01 \AA${}^{-1}$ and using the HERIX spectrometer arm with an ultimate resolution of about 1.5\,meV at the incident photon energy of 23.7\,keV ($\lambda \!=\! 0.5226$\,\AA). The use of an X-ray probe is preferable over neutrons for hydrogenated samples (due to the large incoherent neutron cross-section from H) and is expected to be highly sensitive to the dynamics of the PbI$_3$ inorganic network because of the high electron densities of Pb and I (large atomic numbers). In order to perform measurements in transmission mode and thus gain access to a broader portion of reciprocal space, we selected a crystal thickness around one absorption length, which at the probing photon energy is approximately 100\,$\mu$m for both MAPbI$_3$ and MAPbBr$_3$. However, the size of I-based crystals could not be optimized to the desired thickness, therefore we measured MAPbI$_3$ in reflection mode, a geometry which imposes certain limitations and, unlike for MAPbBr$_3$, prevented the study of reciprocal space cuts along directions (${\hat{u}}_{\mathbf{Q}}$) that selectively probe longitudinal (${\hat{u}}_{\mathbf{Q}} \!\parallel\! \mathbf{Q}$) or transverse (${\hat{u}}_{\mathbf{Q}} \!\perp\! \mathbf{Q}$) modes. The samples have been mounted on a Cu sampleholder and installed on a Displex cryostat with a Be dome for temperature-dependent measurements.
The spectral resolution, or instrumental response function, for the spectrometer used in this study was measured by collecting the scattered signal from a Plexiglass\textsuperscript{\textregistered} slab. The corresponding energy scan is shown in Fig.\,\ref{SI_Resolution} on a linear scale, with the resolution is calculated as the full-width-at-half-maximum (FWHM) and amounting to 1.5\,meV. The experimental resolution profile $R \left( \omega \right)$ has been directly used in the convolution with the model function ${I}_{\mathrm{DHO}} \left( \mathrm{Q}, \omega \right)$ introduced in the main text (Eq.\,1), yielding the generalized fit function:
\begin{equation}
{I}_{\mathrm{fit}} = \int d {\omega}^{\prime} \left[ {I}_{\mathrm{DHO}} \left( \mathrm{Q}, {\omega}^{\prime} \right) \cdot \left( n \left( {\omega}^{\prime} \right) + 1 \right) \right] \times R \left( \omega - {\omega}^{\prime} \right)
\end{equation}
where the Stokes/anti-Stokes asymmetry is accounted for via the factor $\left[ n \left( \omega \right) +1 \right]$ where $ n \left( \omega \right) = {\left( {e}^{\beta \omega} -1 \right)}^{-1} $ is the Bose-Einstein function ($\beta = 1 / KT$).
\begin{figure}[b!]
\includegraphics[width=1\linewidth]{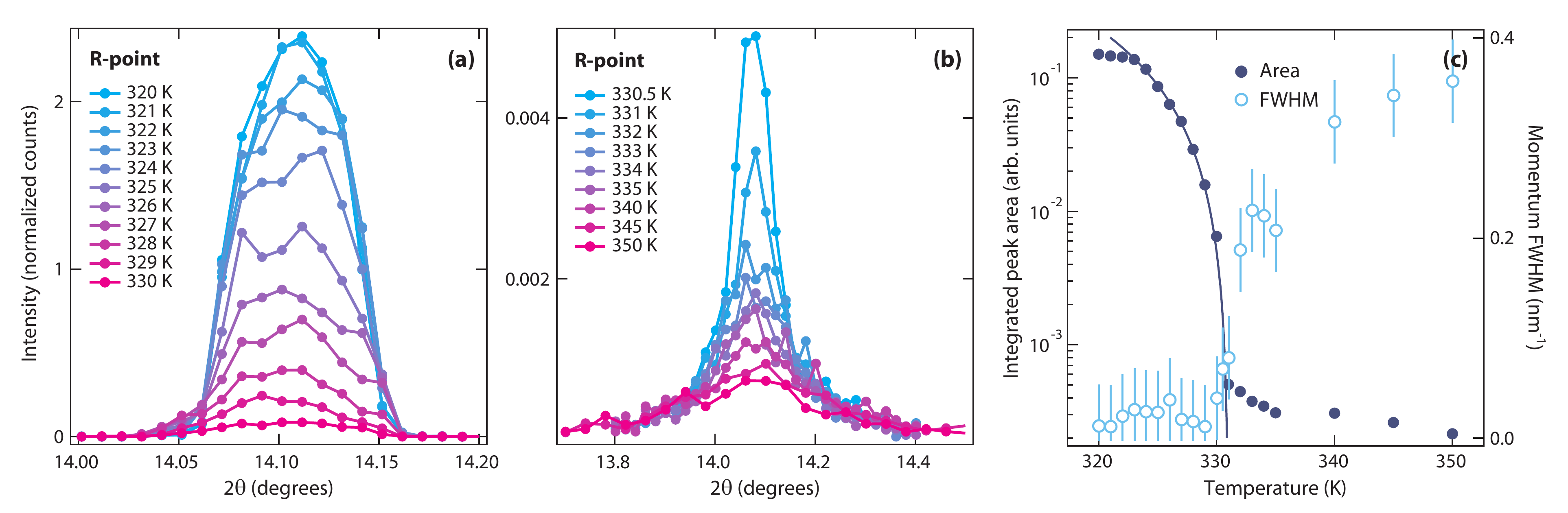}
\caption{(a) Temperature-dependent momentum scans across the $R$-point [${\mathbf{Q}}_{R} \!=\! \left( 3/2, 1/2, 5/2 \right)$] of MAPbI$_3$ below and around the phase transition temperature ${T}_{\mathrm{c}}$ (a) and above ${T}_{\mathrm{c}}$, in the nominally cubic phase (b). Note the very different angular scales in (a) and (b), owing to the short-ranged nature of the remnant tetragonal correlations above ${T}_{\mathrm{c}}$. (c) Corresponding integrated area (dark blue, left axis) and peak FWHM linewidth (light blue, right axis) as a function of temperature.}
\label{SI_Fig1}
\end{figure}

\subsection{Extended momentum- and energy-dependent scans in MAPbI$_3$}

Figure\,\ref{SI_Fig1} shows additional momentum scans of MAPbI$_3$ on a more extended temperature range around the $R$-point [${\mathbf{Q}}_{R} \!=\! \left( 3/2, 1/2, 5/2 \right)$] and across the cubic-to-tetragonal phase transition ${T}_{\mathrm{c}}$. These profiles have been acquired by scanning the detector angle ($2 \theta$) at fixed sample position, and by selecting only the elastic scattering ($\omega \!=\! 0$) portion of the IXS spectrum. The momentum scans below or near ${T}_{\mathrm{c}}$ [Fig.\,\ref{SI_Fig1}(a), also shown in Fig.\,3(a) of the main text] exhibit a lineshape with sharp tails, which represents the projection, at the photon detector, of the exit slits of the HERIX spectrometer, and in general cannot be fitted as a Gaussian or Lorentzian profile. These scans suggest that the intrinsic momentum linewidth is much smaller than the momentum resolution of the spectrometer and are therefore indicative of the presence of long-ranged crystalline order with tetragonal symmetry (the Bragg peak intensity at the $R$-point is a \textit{bona fide} order parameter for the tetragonal structure). Above the phase transition [Fig.\,\ref{SI_Fig1}(b)] the momentum scans become increasingly broader and weaker in amplitude, however a broad Bragg reflection can be still seen at the highest temperature (350\,K), suggesting the persistence of tetragonal domains with short-ranged order well above the thermodynamic transition point. The peak area and momentum linewidth $\Delta Q$ (evaluated as FWHM and dimensioned in ${\mathrm{nm}}^{-1}$) are shown in Fig.\,\ref{SI_Fig1}(c). However we point out that the FWHM should be regarded as an overestimate, or upper boundary, of the real values, due to the tradeoff between momentum resolution and counting statistics. In any case, and besides the presence of a clear singularity at ${T}_{\mathrm{c}}$ for both the peak area and FWHM, we observe that tetragonal nanopuddles are still present up to 20\,K above the transition temperature, with an average size or local coherence length of at least $\xi = {\Delta Q}^{-1} \sim 3$\,nm. Furthermore, the temperature dependence suggest an asymptotic behavior for the FWHM, which seems to stabilize around an average length scale, possibly reflecting the spatial extent of the local perturbation of cubic symmetry around a symmetry-breaking defect.
\begin{figure}[t!]
\includegraphics[width=0.7\linewidth]{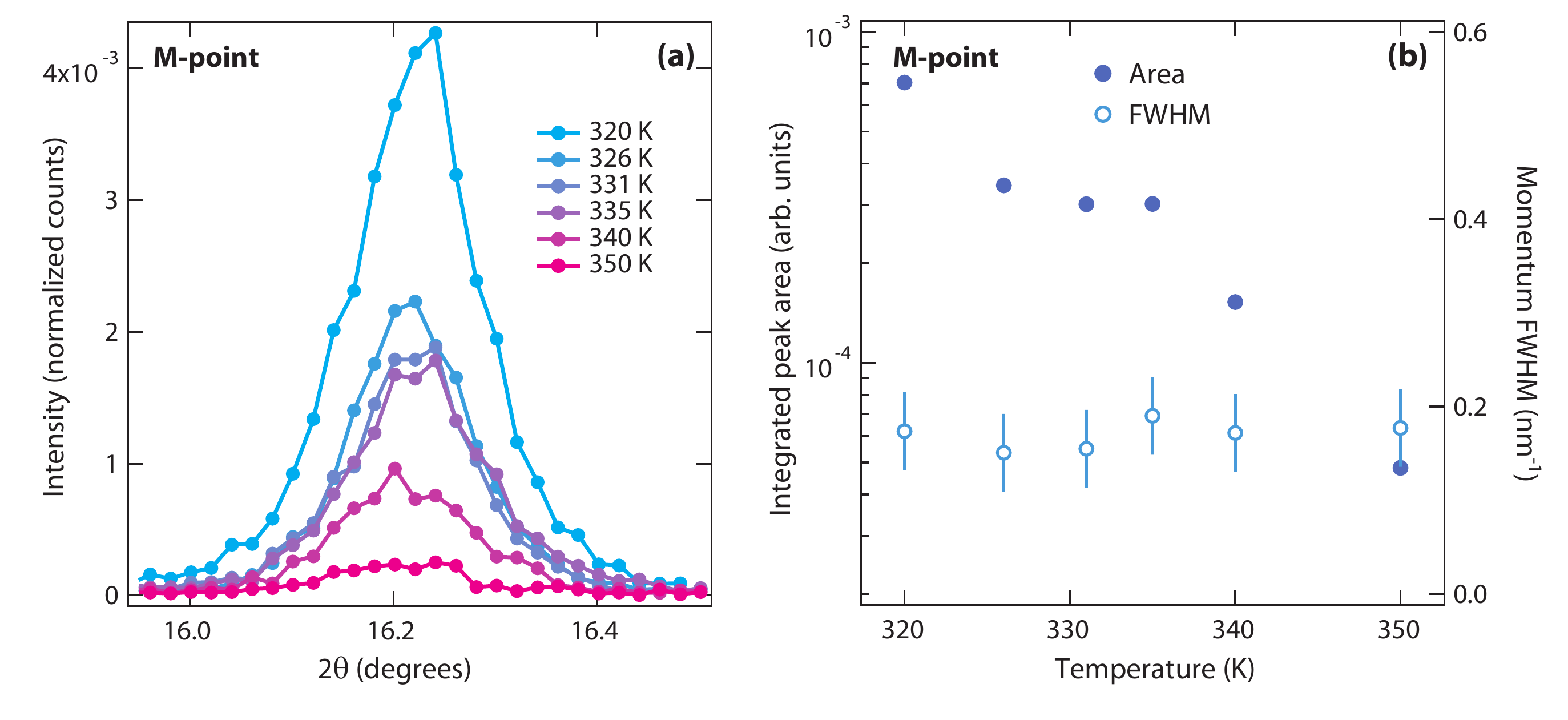}
\caption{(a) Temperature-dependent momentum scans around the $M$-point [${\mathbf{Q}}_{M} \!=\! \left( 3/2, 1/2, 3 \right)$] of MAPbI$_3$ across the phase transition temperature ${T}_{\mathrm{c}}$. (b) Corresponding integrated area (dark blue, left axis) and peak FWHM linewidth (light blue, right axis) as a function of temperature.}
\label{SI_Fig2}
\end{figure}
\begin{figure}[b!]
\includegraphics[width=0.6\linewidth]{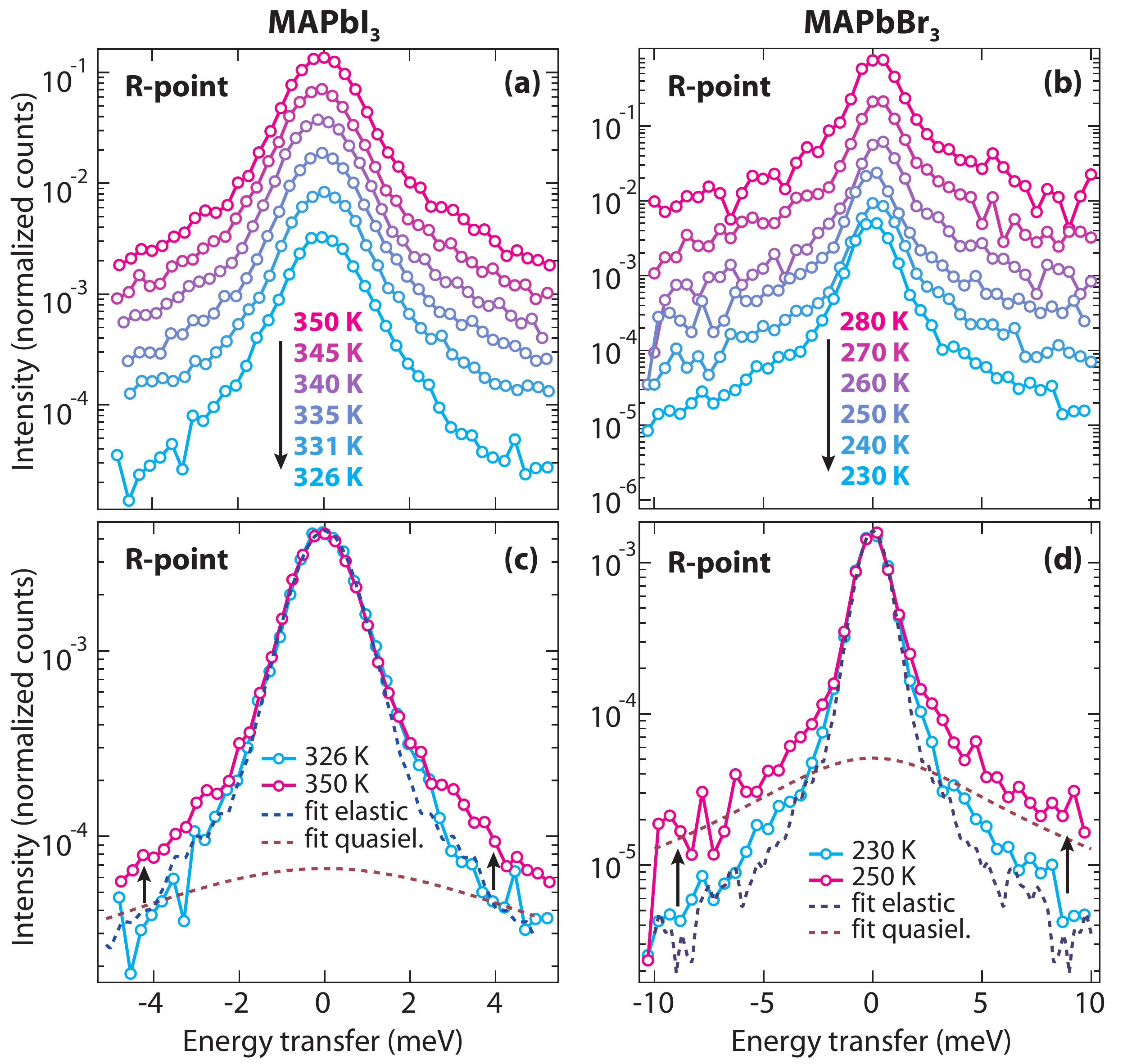}
\caption{(a,b) Series of inelastic scans at the $R$-point across the cubic-to-tetragonal phase transition in MAPbI$_3$ (a) and MAPbBr$_3$ (b). (c,d) Detailed comparison of low- and high-temperature scans in MAPbI$_3$ (c) and MAPbBr$_3$ (d), highlighting the emergence of spectral weight at finite energy transfer in the cubic phase, revealing a regime of nanoscale dynamical fluctuations and coexisting structural orders even well above the phase transition.}
\label{SI_Fig3}
\end{figure}
Momentum scans at the $M$-point [${\mathbf{Q}}_{M} \!=\! \left( 3/2, 1/2, 3 \right)$] have been measured under the same conditions as for the $R$-point, and are plotted in Fig.\,\ref{SI_Fig2}(a) for a few temperature values across ${T}_{\mathrm{c}}$. While no scattered intensity should be expected at this location in reciprocal space (it is a forbidden reflection for the tetragonal space group $I 4 / m c m$), we observe a weak but detectable peak which is however very  broad [see peak area and momentum linewidth in Fig.\,\ref{SI_Fig2}(b)]. We interpret this finding with the possibility that small nanoregions breaking locally breaking tetragonal symmetry are nucleated around defects and persist over an extended temperature range. Since the spontaneous symmetry lowering of the cubic structure favours a transition to a tetragonal symmetry at ${T}_{\mathrm{c}}$, these local nanoregions deviating from tetragonal order never become long-ranged, as they do not represent a generalized instability of the lattice (unlike at $R$-point, where phonon condensation occurs), thereby explaining the temperature-independent FWHM as shown in Fig.\,\ref{SI_Fig2}(b).

We further explored the energy-dependence of the low-energy scattering around the $R$-point, across the phase transition. The corresponding IXS scans data are shown in Figs.\,\ref{SI_Fig3}(a) and (b) for MAPbI$_3$ and MAPbBr$_3$, respectively, with corresponding low- and high-temperature profiles in Figs.\,\ref{SI_Fig3}(c) and (d). These scans are fitted with an elastic component (resolution-limited) and a quasielastic component (with finite broadening). We find that the in the low-temperature scans the scattering is purely elastic (i.e., arising from static order, possibly quenched by local defect), while the high-temperature scans exhibit a quasielastic component with an energy scale of a few meV which could reflect the presence of an overdamped soft phonon, or dynamical domain fluctuations with a corresponding timescale of order ps. In any case, our temperature-dependent scattering seem to confirm the tendency of the lattice to deviate from its idealized structure with sub-ps timescale as proposed by Quarti \textit{et al.} \cite{quarti_interplay_2014}. However, we cannot determine whether these dynamical effects simply correspond to overdamped lattice vibrations at the $R$-point or to some other mechanism.

Lastly, we show the quasielastic IXS scans at the $M$-point in Fig.\,\ref{SI_Fig3}(a), again for selected temperatures spanning the cubic-to-tetragonal phase transition. At a first inspection, no ostensible changes would be apparent, however when contrasting the high- to low-temperature scans [Fig.\,\ref{SI_Fig3}(b)] we notice (similar to the case of the $R$-point discussed in the main text) an increase in the quasielastic intensity on top of the main (resolution-limited) elastic line. This quasielastic intensity, which is manifested in the appearance of additional spectral weight on the tails of the peak around 5\,meV, may reflect the increase in phonon population or it might be indicative of dynamical processes involving the breaking of tetragonal symmetry (thereby producing scattered intensity at the $M$-point) on a timescale of $\hbar / 5\,\mathrm{meV} \sim 0.8$\,ps.
\begin{figure}[t!]
\includegraphics[width=0.6\linewidth]{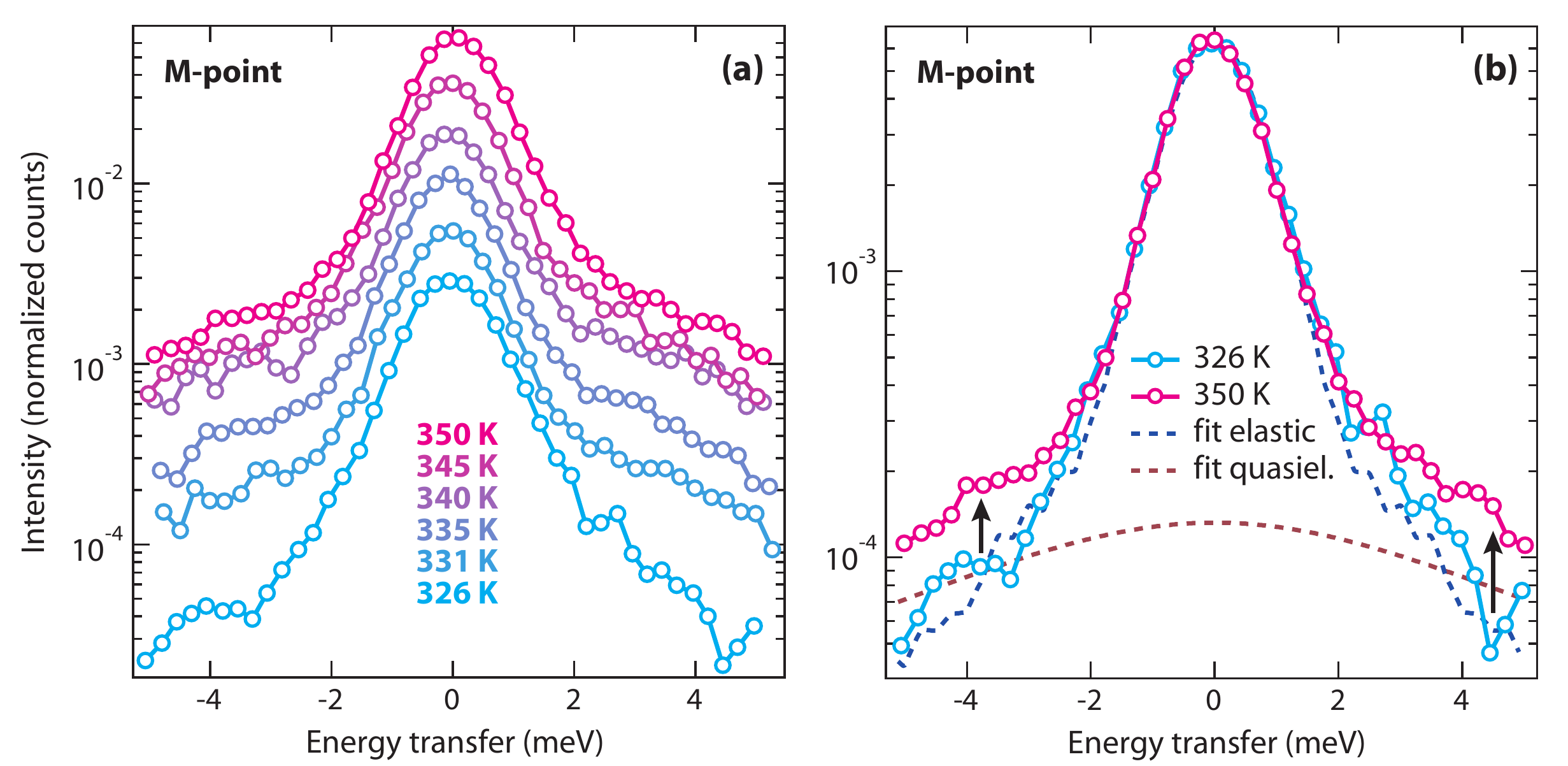}
\caption{(a) Temperature-dependent IXS scans at the $M$-point of MAPbI$_3$ across the phase transition temperature ${T}_{\mathrm{c}}$. (b) Selected view of the highest- and lowest-temperature scans, showing the appearance of inelastic spectral weight for the scan above ${T}_{\mathrm{c}}$.}
\label{SI_Fig4}
\end{figure}

\subsection{X-ray Powder Diffraction}

X-ray powder diffraction measurements were performed on the bending magnet station at DND-CAT sector 5 of the Advanced Photon Source. The X-ray wavelength used was 0.4\,\AA, selected to reduce X-ray absorption by the sample. The sample data were collected in flat plate geometry, with the Cu plate installed in an Advanced Research Systems cryostat and collected over a temperature range of 12\,K to 350\,K. Data were collected using a 1-D solid state Cyberstar detector in Bragg-Brentano geometry. Typical step sizes in two-theta were 0.0015\,deg.  

\subsection{Single-Crystal Neutron Diffraction}

Variable temperature experiment for MAPbI$_3$ was performed using the TOPAZ single crystal diffractometer at the Spallation Neutron Source, Oak Ridge National Laboratory \cite{Jogl_2011}. Sample temperature control in the range of 350\,K -- 190\,K used Oxford Cryosystems’ Cryostream Plus 700 with a LN$_2$-gasflow setup.  A block-shaped deuterated crystal of d$_6$-MAPbI$_3$ with the dimensions of $0.86 \times 1.0 \times 1.5 $ mm$^3$ was mounted onto the tip of a MiTeGen loop with Super Glue, and transferred to the TOPAZ goniometer for data collection in neutron wavelength-resolved Laue mode, in which a continuous 3D volume of reciprocal space can be measured from a stationary crystal.  Using the initial orientation matrix obtained at room temperature, the $\left( 3/2,1/2,1/2 \right)$ superlattice peak in the cubic $Pm\bar{3}m$ cell was placed on a forward scattering detector with the aid of CrystalPlan software \cite{Zikovsky_2011}. The single crystal sample was then heated to 350\,K at a ramp rate of 1\,K/min. The super lattice peak of the cubic phase dissipated at 350\,K, evidence by the very low $I/\sigma \left( I \right)$ ratio. The sample was then cooled stepwise from 350\,K to 340, 335, 330, 325, 320, 315, 310, 305, 300, 295, 270, 250, 230, 190\,K, respectively. Single crystal neutron diffraction pattern at each temperature was measured for approximately 2 hrs. Integrated raw Bragg intensities were obtained using the 3-D ellipsoidal Q-space integration method \cite{Schultz_TOPAZ}. Refinement of the lattice parameter at each temperature used intensity data with $I/\sigma \left( I \right) > 5$ to high resolution shells with ${d}_{\mathrm{min}}$ of 0.6\,\AA. Data reduction, including neutron TOF spectrum, detector efficiency, Lorentz and absorption corrections, was carried out with the ANVRED3 program \cite{Schultz_1984}. Table\,\ref{Neutron_data} lists the normalized Bragg peak intensities for the $\left( 3/2,1/2,1/2 \right)$ super lattice peak and corresponding $(c-a)/c$ ratios.

\begin{table}[t!]
\begin{center}
\begin{tabularx}{0.6\textwidth}{c *{4}{Y}}
\toprule
\toprule
$T$ (K) & Intensity, $I$ & ${\sigma}_{I}$ & $ \left( c-a \right) / c $ & ${\sigma}_{\left( c-a \right) / c} $  \\
 & & & & \\
\midrule
\midrule
190 & 217.1 & 25.7 & 0.01805 & 0.00014 \\
210 & 203.8 & 24.9 & 0.01676 & 0.00014 \\
230 & 187.8 & 23.8 & 0.01482 & 0.00015 \\
250 & 174.9 & 23.0 & 0.01332 & 0.00017 \\
270 & 146.6 & 21.0 & 0.01136 & 0.00016 \\
295 & 117.6 & 18.5 & 0.01015 & 0.00017 \\
300 & 108.6 & 18.0 & 0.01000 & 0.00011 \\
305 & 100.5 & 17.3 & 0.00686 & 0.00016 \\
310 & 84.7 & 15.9 & 0.00762 & 0.00016 \\
315 & 72.7 & 14.6 & 0.00636 & 0.00021 \\
320 & 58.1 & 13.1 & 0.00463 & 0.00019 \\
325 & 14.5 & 6.5 & 0.00081 & 0.00014 \\
330 & 2.8 & 5.0 & 0.00167 & 0.00019 \\
335 & 11.9 & 5.7 & 0.00078 & 0.00014 \\
340 & 2.0 & 5.3 & 0.00033 & 0.00018 \\
350 & 0.7 & 5.5 & 0.00086 & 0.00019 \\
\bottomrule
\bottomrule
\end{tabularx}
\end{center}
\caption{Bragg peak intensity and (c-a)/c ratio from neutron single crystal diffraction.}
\label{Neutron_data}
\end{table}


\begin{thebibliography}{35}
\bibitem [{HKL()}]{HKL_note}%
  \BibitemOpen
  \href@noop {} {}\bibinfo {note} {{Here, $H$, $K$, and $L$ are integer
  indices, which are also used to label the reciprocal lattice location where
  measurements were perfomed}}\BibitemShut {NoStop}%
\end{thebibliography}
\end{document}